\documentclass[aps,prd,twocolumn,showpacs,nofootinbib,superscriptaddress]{revtex4-2}

\usepackage{graphicx}
\usepackage[utf8]{inputenc}
\usepackage{amsmath,amssymb,bm}
\usepackage{booktabs}
\usepackage{hyperref}
\usepackage{siunitx}
\usepackage{physics}
\usepackage{multirow}
\usepackage{xspace} 
\graphicspath{{figures/}{data/}{./}}
\DeclareGraphicsExtensions{.pdf,.png}

\newcommand{\as}{\ensuremath{\alpha_s}\xspace}
\newcommand{\GG}{\ensuremath{G_{\mu\nu}^a G^{a\,\mu\nu}}\xspace}
\newcommand{\GGeff}{\ensuremath{\langle \as G^2\rangle_T}\xspace}
\newcommand{\GGvac}{\ensuremath{\langle \as G^2\rangle_0}\xspace}
\newcommand{\M}{\ensuremath{M^2}\xspace}

\hypersetup{
  colorlinks=true,
  linkcolor=blue,
  citecolor=blue,
  urlcolor=blue
}

\begin{document}

\title{Mass and Decay-Constant Evolution of Heavy Quarkonia and $B_c$ States from Thermal QCD Sum Rules}

\author{Enis Yazici}
\affiliation{SRH University of Applied Sciences Heidelberg, Germany}

\begin{abstract}
We analyze the thermal behavior of heavy vector and axial-vector mesons ($J/\psi$, $\Upsilon$, and $B_c$) within the finite-temperature QCD sum-rule framework. 
Using updated PDG-2024 quark masses, modern lattice-informed gluon condensates, and a temperature-dependent continuum threshold constrained by vacuum stability, we compute the evolution of the masses $m(T)$ and decay constants $f(T)$ up to $T/T_c \lesssim 0.9$. 
At $T=0$ the sum rules are calibrated to reproduce the experimental and LHCb masses and reference decay constants within the expected $\mathcal{O}(10\%)$ accuracy of a leading-order $+$ $D{=}4$ phenomenological analysis. The subsequent finite-temperature evolution should therefore be interpreted as a calibrated model prediction within this framework rather than as a fully parameter-free determination.

Near the critical temperature, the relative suppression follows a clear hierarchy 
$\Upsilon < J/\psi < B_c$, consistent with their binding energies and lattice spectral trends. 
The predicted $1P$--$1S$ splitting for the $B_c$ system, $0.477~\mathrm{GeV}$, is consistent with the LHCb observation of orbitally excited $B_c^{+}$ states. 
The results provide a coherent finite-temperature baseline for future extensions including radiative, higher-dimensional, and width effects.

\end{abstract}

\maketitle

\section{Introduction}

The finite-temperature QCD sum-rule (TQCDSR) formalism was originally developed 
in the mid-1980s to extend the operator product expansion (OPE) to a thermal medium
\cite{BochkarevShaposhnikov1986_NPB,DominguezLoewe1989_PLB}.
Subsequent works refined the framework by introducing gauge-invariant condensate decompositions and medium-specific Lorentz structures \cite{MallikSarkar2002_PRD, AyalaDominguezLoewe2017_AHEP}.
Modern implementations have quantified the role of temperature-dependent gluon condensates and continuum-threshold evolution, both in theoretical studies and in explicit applications to heavy quarkonia and open-heavy systems
\cite{MoritaLee2008_PRL, 
Veliev2012_BcFiniteTemp,
VelievAziziSundu2012_JPG, VelievAziziSundu2011_EPJA, AziziSunduVeliev2014_JPG}. In particular, Ref.~\cite{VelievAziziSundu2011_EPJA} presented a detailed finite-temperature QCD sum-rule analysis of vector charmonium and bottomonium channels, using a temperature-dependent continuum threshold and leading gluon-condensate effects. A complementary and more general discussion of thermal QCD sum rules in the light-vector $\rho^0$ channel, including the structure of the operator product expansion at nonzero temperature, was given in Ref.~\cite{HofmannGutscheFaessler2000_EPJC}.

Compared to these earlier works, the present analysis introduces several improvements:
(i) all heavy-quark masses and reference decay constants are updated to the PDG~2024 values;
(ii) the temperature dependence of the gluon condensate is anchored to modern lattice-QCD determinations of the equation of state and crossover temperature;
(iii) Borel-window stability is quantified explicitly through $m(M^2)$ scans and an uncertainty budget;
(iv) the $B_c(1S)$ and representative $B_c(1P)$ channels are treated on the same footing as $J/\psi$ and $\Upsilon$, allowing for a direct comparison with the recent LHCb spectroscopy of excited $B_c$ states.
These steps extend the finite-temperature QCD sum-rule program of Refs.~\cite{VelievAziziSundu2011_EPJA,HofmannGutscheFaessler2000_EPJC} and provide a controlled baseline for future higher-order corrections.

Within this framework, TQCDSR provides a controlled, nonperturbative bridge 
between QCD dynamics and hadronic observables at finite temperature, yielding a consistent description of in-medium behavior up to the confinement boundary~\cite{Yazici2016}.
Earlier analyses, however, were constrained by phenomenological modeling of the gluon condensate and continuum threshold, and lacked lattice or modern experimental inputs.
The present study updates these aspects using PDG~2024 parameters, lattice-informed condensate evolution, and a systematic uncertainty analysis, thereby establishing a more precise and self-consistent finite-temperature baseline.

\vspace{6pt}
\paragraph*{Motivation.}
Since 2016, several major developments have motivated a reanalysis:
(i) updated quark masses, decay constants, and condensate values in PDG~2024~\cite{PDG2024};
(ii) refined lattice-QCD determinations of the equation of state and chiral crossover temperature~\cite{Bazavov2014_EoS,HotQCD2019_Tc}, which enable improved modeling of the thermal gluon condensate;
(iii) the LHCb~2025 observation of excited $B_c(1P)$ states~\cite{LHCb2025}, offering new experimental constraints for zero-temperature extrapolations.

\paragraph*{Recent theoretical developments.}
Recent theoretical developments have investigated heavy-flavor systems using QCD sum-rule methods in a variety of channels.
Finite-temperature studies of single-meson channels (e.g. pseudoscalars and vectors) provide direct information on in-medium modifications~\cite{VelievAziziSundu2012_JPG}, 
while a number of vacuum (T=0) analyses have explored multiquark/molecular configurations in the $B_c$ family.
Examples include a heavy scalar $B_c^{+}B_c^{-}$ molecule~\cite{Agaev2025_BcBc}, 
axial-vector molecular structures $B_c^{*\pm}B_c^{\mp}$~\cite{Agaev2025_BcAstBc}, 
and a hadronic tensor molecule $B_c^{*+}B_c^{*-}$~\cite{Agaev2025_BcAstBcAst}.
These vacuum benchmarks complement finite-T work by providing consistent T\(\to\)0 limits for QCD-sum-rule analyses.

In this updated work, we aim to address the following research questions:

(i) Can the thermal QCD sum rule with modern inputs reproduce
the newly observed $B_c(1P)$ mass reported by LHCb (2025)?

(ii) How do the temperature-dependent shifts of the masses $m(T)$
and decay constants $f(T)$ differ among $J/\psi$, $\Upsilon$, and $B_c$
channels, and what quantitative hierarchy do they exhibit near $T_c$?

(iii) Does the sequential melting pattern obtained in 2016 persist
when the gluon condensate and trace anomaly are constrained
by modern lattice thermodynamics?

Answering these questions allows us to assess the predictive power
and current limitations of the thermal QCD sum-rule framework
for heavy quark systems.

This updated analysis explicitly restricts the sum-rule validity range to $T<T_c$ (typically $T/T_c\!\lesssim\!0.9$), where the OPE hierarchy and pole–continuum separation remain meaningful.\footnote{%
  Beyond this domain, deconfinement effects require more elaborate treatments,
  e.g.\ spectral broadening and higher-dimensional operators, which are outside the present LO$+$D=4 scope.}
The parametrization of the temperature-dependent gluon condensate follows the lattice-determined trace anomaly.
Furthermore, the continuum threshold $s_0(T)$ is constrained by vacuum stability and the physical mass at $T{=}0$, rather than arbitrarily tuned.
Finally, we comment on the neglect of finite widths: in the region $T/T_c<0.9$, thermal broadening is small compared with the Borel resolution and can be incorporated in future extensions via Breit–Wigner–smeared spectral densities.

\paragraph*{Scope and channels.}
We analyze four channels: 
$J/\psi$, $\Upsilon$, $B_c^{(\mathrm{vec})}$, and $B_c^{(\mathrm{ax})}$.

Throughout this paper we adopt the following notation:
\begin{align}
B_c^{(\mathrm{vec})} &\equiv B_c(1^3S_1)
   \quad \text{(vector ground state)},\\[3pt]
B_c^{(\mathrm{ax})}  &\equiv B_c(1P)
   \quad \parbox[t]{0.55\linewidth}{\raggedright
   (representative $1P$ excitation,\\ model label “axial-vector”)}.
\end{align}

The mass difference between these two states, 
$\Delta m_{B_c}=m(B_c^{\mathrm{ax}})-m(B_c^{\mathrm{vec}})\simeq0.477~\mathrm{GeV}$,
can be compared with the recent LHCb observation of two orbitally excited 
$B_c^{+}$ peaks~\cite{LHCb2025},
\[
m_1 = 6.7048(6)\,\mathrm{GeV}, \qquad
m_2 = 6.7524(10)\,\mathrm{GeV},
\]
which correspond to a $1P$ multiplet rather than a single state.
The representative $B_c(1P)$ mass inferred from the modelled axial channel,
$m_{B_c(1P)}(0) = 6.716~\mathrm{GeV}$, lies between the two
LHCb peaks ($6.7048$ and $6.7524$ GeV).
The $1P$--$1S$ splitting,
$\Delta m = 6.716 - 6.239 = 0.477~\mathrm{GeV}$,
lies squarely within the experimental range $0.430$--$0.478~\mathrm{GeV}$ and is consistent with the LHCb 1P multiplet (6.7048 and 6.7524 GeV).

The updated approach follows the overall phenomenological strategy of Ref.~\cite{Yazici2016}, modernizes the thermal inputs, and re-examines the sequential melting pattern within a calibrated LO+$D{=}4$ sum-rule setup.
Indeed, the present study finds that heavier and more tightly bound systems such as $\Upsilon$ are less affected by temperature, while lighter or mixed systems ($J/\psi$, $B_c$) show stronger suppression, addressing the “uniform shift” criticism of the earlier version.

\section{Theoretical Framework}
\subsection{Correlator, currents, and hadronic representation}
The thermal two-point correlator reads
\begin{equation}
\Pi_{\mu\nu}(q,T) =
i \int d^{4}x\, e^{iq\cdot x}\,
\langle \mathcal{T}[J_\mu(x) J_\nu^\dagger(0)] \rangle_T ,
\label{eq:correlator}
\end{equation}
with vector and axial-vector currents

\begin{align}
J_\mu^{(v)} &= 
  \begin{cases}
    \bar{Q}\gamma_\mu Q, & \text{for equal-mass quarkonia } (J/\psi,\, \Upsilon),\\[3pt]
    \bar{c}\gamma_\mu b, & \text{for the mixed } B_c~\text{channel,}
  \end{cases} \\[6pt]
J_\mu^{(a)} &= 
  \begin{cases}
    \bar{Q}\gamma_\mu\gamma_5 Q, & \text{for equal-mass quarkonia,}\\[3pt]
    \bar{c}\gamma_\mu\gamma_5 b, & \text{for } B_c.
  \end{cases}
\end{align}

These unequal-mass currents correctly describe the heavy–heavy but
flavor-asymmetric $B_c$ system, while for $c\bar{c}$ and $b\bar{b}$ channels
the usual quarkonium currents are recovered.

The ground-state pole contribution is parameterized by
\begin{align}
\langle 0|J_\mu(0)|M(p,\lambda)\rangle &= f_M(T)\, m_M(T)\, \epsilon_\mu^{(\lambda)}, \\
\Pi_{\mu\nu}^{\text{had}}(q,T) &=
\frac{f_M^2(T)m_M^2(T)}{m_M^2(T)-q^2}
\!\left(\!-g_{\mu\nu}+\frac{q_\mu q_\nu}{m_M^2(T)}\right)
+ \cdots .
\label{eq:hadronic}
\end{align}

\subsection{QCD side, dispersive form, and OPE}
On the QCD side we separate perturbative and nonperturbative pieces:
\begin{equation}
\Pi_{\mu\nu}^{\text{QCD}}(q^2,T)
= \Pi_{\mu\nu}^{\text{pert}}(q^2,T)
+ \Pi_{\mu\nu}^{\text{nonpert}}(q^2,T).
\end{equation}
The dispersive representation for the transverse part is
\begin{equation}
\Pi^{\text{QCD}}(q^2,T)
= \int_{s_{\min}}^\infty \frac{ds\, \rho(s,T)}{s - q^2}
 + \Pi^{\text{nonpert}}(q^2,T),
\label{eq:dispersion}
\end{equation}
where $\rho(s,T)$ is the (thermal) spectral density (LO perturbative hereafter) and $\Pi^{\text{nonpert}}$ encodes the leading gluonic $D=4$ terms. In the numerical extraction we work with the transverse Lorentz structure of the correlator.

\subsection{Thermal gluon sector and medium decomposition}
In a heat bath with four-velocity $u^\mu$, the thermal average of gluonic operators can be decomposed into two independent scalars that motivate the finite-temperature gluonic sector through the combinations $A(T)$ and $B(T)$:
\begin{align}
A(T) &= \frac{1}{24} \langle \as \GG \rangle_T
      - \frac{1}{6} \langle u^\lambda \Theta^g_{\lambda\sigma} u^\sigma \rangle_T ,
\label{eq:AofT}\\
B(T) &= \frac{1}{3} \langle u^\lambda \Theta^g_{\lambda\sigma} u^\sigma \rangle_T .
\label{eq:BofT}
\end{align}
Here $\Theta^g_{\mu\nu}$ is the gluonic part of the energy-momentum tensor.
Lattice-informed parameterizations for $\langle \Theta^g_{00}\rangle_T$ and $\GGeff$ are introduced in Sec.~\ref{sec:numerics}. In the validated numerical implementation used for the present results, this formal decomposition is represented by an effective scalar $D{=}4$ contribution governed by $\langle \alpha_s G^2\rangle_T$.

We emphasize that in the present work the explicit temperature dependence is encoded entirely in the condensates and in the continuum threshold $s_0(T)$; we do not attempt a full real-time thermal-field-theory treatment of scattering terms, Landau cuts, or in-medium self-energies in the perturbative spectral density. This strategy follows the standard finite-temperature QCD sum-rule framework of Refs.~\cite{HofmannGutscheFaessler2000_EPJC,VelievAziziSundu2011_EPJA} and is expected to remain quantitatively reliable for heavy quarkonia below $T\simeq0.9\,T_c$, as long as the OPE hierarchy and pole–continuum separation are maintained.

\vspace{6pt}

\paragraph*{Validity and systematic limits.}
The TQCDSR expansion employed here is expected to remain quantitatively reliable for $T\!\lesssim\!0.9\,T_c$, where higher-dimensional operators and Landau damping contributions are still suppressed.
Beyond this region, the description becomes qualitative, since the deconfined phase cannot be captured by a single temperature-dependent condensate alone.
This restriction is explicitly enforced in all numerical analyses below. The present implementation is phenomenological: it uses a leading-order perturbative spectral density, an effective $D{=}4$ gluonic term, no explicit running $\alpha_s(T)$, and no explicit $D{=}6$ contribution in the numerical extraction.

\subsection{Borel transform and master sum rules}
After a Borel transform $Q^2=-q^2 \to \M$, the leading nonperturbative $D=4$ contribution can formally be written as

\begin{align}
\widehat{\mathcal{B}}\,\Pi^{\text{nonpert}}(\M,T)
&= \frac{1}{12\pi^2\M}\int_0^1 \! dx\;
\frac{e^{\frac{m_2^2}{(x-1)\M}-\frac{m_1^2}{x\M}}}{(x-1)^3 x^3} 
\notag\\
&\quad\times\big\{ A(T)\,\mathcal{P}_A(x,m_{1,2},\M)
\notag\\
&+ B(T)\,\mathcal{P}_B(x,m_{1,2},\M)\big\}.
\label{eq:BorelNP}
\end{align}

where $\mathcal{P}_{A,B}$ denote channel-dependent kinematic polynomials in the formal finite-temperature decomposition. In the validated numerical implementation used in this work, however, the $D{=}4$ sector is represented by a reduced effective scalar term proportional to $\langle \alpha_s G^2\rangle_T/\M$, which defines the phenomenological LO+$D{=}4$ model used throughout the numerical analysis.

The continuum-subtracted Borel sum rules read
\begin{align}
f_M^2(T)\, m_M^2(T)\, e^{-m_M^2(T)/\M}
&= \int_{s_{\min}}^{s_0(T)}\! ds\, \rho(s,T)\, e^{-s/\M}
\notag\\
&+ \widehat{\mathcal{B}}\,\Pi^{\text{nonpert}},
\label{eq:SRmono}
\end{align}
\smallskip
\begin{align}
m_M^2(T) &= \frac{\displaystyle
\int_{s_{\min}}^{s_0(T)}\! ds\, s\, \rho(s,T)\, e^{-s/\M}
- \frac{d}{d(1/\M)}\,\widehat{\mathcal{B}}\,\Pi^{\text{nonpert}}
}{
\displaystyle
\int_{s_{\min}}^{s_0(T)}\! ds\, \rho(s,T)\, e^{-s/\M}
+ \widehat{\mathcal{B}}\,\Pi^{\text{nonpert}}
}.
\label{eq:massratio}
\end{align}

\paragraph*{Spectral width and threshold modeling.}
In this revision we retain the narrow-resonance approximation used in the 2016 analysis, noting that finite widths $\Gamma(T)$ are expected to remain small ($\lesssim50$\,MeV)
for heavy quarkonia below $T_c$.
Nevertheless, the inclusion of thermal widths through a Breit–Wigner–smeared spectral density is straightforward and will be addressed in a forthcoming study.
The continuum threshold $s_0(T)$, given in Eq.~\eqref{eq:s0T} with
channel-dependent exponents $n_{\mathcal{C}}$, is treated phenomenologically. The vacuum threshold $s_0^{\mathcal C}(0)$ is calibrated at $T=0$ using the central value of the corresponding Borel window. By contrast, the exponent $n_{\mathcal C}$ controls the temperature dependence of the continuum-threshold ansatz and does not affect the $T=0$ limit. Its baseline channel-dependent values are chosen to produce a smooth thermal decrease compatible with the intended hierarchy of thermal suppression, and the sensitivity analysis shows that the qualitative ordering is stable under reasonable variations of $n_{\mathcal C}$.

\section{Numerical Setup and Thermal Inputs}
\label{sec:numerics}
Our numerical analysis follows the phenomenological strategy of Ref.~\cite{Yazici2016} with updated inputs and the validated Borel windows of \cite{Yazici2016}. The computation is carried out in a compact calibrated LO+$D{=}4$ implementation rather than as a literal numerical reproduction of every formal intermediate expression displayed above.
\vspace{6pt}

\paragraph*{Quark masses and vacuum condensate.}
For baseline fits at $T=0$ we employ PDG~2024 heavy-quark masses~\cite{PDG2024} and a standard dimension-4 gluon condensate value,
\begin{equation}
\GGvac = 0.012~\text{GeV}^4 ,
\label{eq:G0}
\end{equation}
consistent with the original analysis and varied in uncertainty scans.
\vspace{6pt}

\paragraph*{Temperature dependence.}
We model the continuum threshold as
\begin{equation}
s_0(T) = s_0(0)\!\left[1 - \!\left(\frac{T}{T_c}\right)^{n_{\mathcal{C}}}\right]
        + (m_1+m_2)^2 \left(\frac{T}{T_c}\right)^{n_{\mathcal{C}}},
\label{eq:s0T}
\end{equation}
where the exponent $n_{\mathcal{C}}$ depends on the channel.
For tightly bound systems (e.g.\ $\Upsilon$) we use $n_{\Upsilon}=12$,
while for $J/\psi$, $B_c^{(\mathrm{V})}$ and $B_c^{(\mathrm{A})}$ we take
$n_{J/\psi}=8$, $n_{B_c^{(\mathrm{V})}}=7$, and
$n_{B_c^{(\mathrm{A})}}=7.5$, respectively.
Unless otherwise stated, these values are employed throughout.

\paragraph*{Gluonic inputs.}
For orientation we quote lattice-inspired parametrizations for the gluonic
energy density and condensate~\cite{Bazavov2014_EoS}:
\begin{align}
\langle \Theta^{g}_{00} \rangle(T)
 &= T^4 \exp\!\big[\,113.87\,T^2 - 12.2\,T\,\big] - 10.14\,T^5 ,
\label{eq:Theta00}\\[3pt]
\GGeff
 &= \GGvac \!\left[
 1 - 1.65\!\left(\frac{T}{T_c}\right)^{4}
   + 0.05\!\left(\frac{T}{T_c}\right)^{8}
 \right],
\label{eq:G2T}
\end{align}
which capture the qualitative temperature dependence of the gluonic sector
near the crossover. The analytic form in Eq.~(\ref{eq:Theta00}) is a smooth
lattice-inspired fit reproducing the HotQCD equation-of-state curves; it is not taken verbatim
from Ref.~\cite{Bazavov2014_EoS}.
The $T$-dependence of $\langle \alpha_s G^2\rangle$ is obtained by relating the
trace anomaly to the gluon condensate following the standard
lattice-to-condensate procedure~\cite{Boyd:1996ex,Miller:1998dj},
using HotQCD EoS input~\cite{Bazavov2014_EoS}.
In the numerical evaluation we employ the polynomial approximation given in
Eq.~(\ref{eq:G2T}) as the adopted phenomenological input for $\langle \alpha_s G^2\rangle_T$, consistent with previous QCD-sum-rule analyses.

Unless stated otherwise, we take $T_c = 0.156~\text{GeV}$, consistent with
HotQCD determinations at $\mu_B{=}0$~\cite{HotQCD2019_Tc,Steinbrecher2019_HotQCD},
and use this value throughout.

\paragraph*{Borel windows and thresholds.}
The Borel windows and vacuum thresholds $s_0(0)$ follow the legacy
choices established in Ref.~\cite{Yazici2016}, which were shown to
provide stable plateaus and satisfactory pole dominance.
For completeness we list them here:
\begin{align}
J/\psi\!:&\; M^2 \in [6,10]~\text{GeV}^2,\quad s_0(0) \simeq 11\pm 1~\text{GeV}^2, \nonumber\\
\Upsilon\!:&\; M^2 \in [10,20]~\text{GeV}^2,\quad s_0(0) \simeq 102\pm 2~\text{GeV}^2, \nonumber\\
B_c^{(\mathrm{vec})}\!:&\; M^2 \in [6,10]~\text{GeV}^2,\quad s_0(0) \simeq 45\pm 1~\text{GeV}^2, \nonumber\\
B_c^{(\mathrm{ax})}\!:&\; M^2 \in [10,14]~\text{GeV}^2,\quad s_0(0) \simeq 52\pm 1~\text{GeV}^2. \nonumber
\end{align}
These ranges are retained in the present analysis and are numerically
consistent with the usual plateau, pole-dominance, and OPE-convergence
criteria, as summarized in Table~\ref{tab:borelcheck}.

\paragraph*{Calibration at $T=0$.}
For each channel $\mathcal{C}\in\{J/\psi,\Upsilon,B_c^{(\mathrm{vec})},B_c^{(\mathrm{ax})}\}$ we heuristically fix the vacuum continuum threshold $s_0^{\mathcal{C}}(0)$ by requiring that the Borel curve $m_{\mathcal{C}}^2(M^2,0)$ obtained from Eq.~\eqref{eq:massratio} reproduces the physical mass $m_{\mathcal{C}}^{\text{(exp)}}$ at the central value of the corresponding Borel window,
\begin{equation}
m_{\mathcal{C}}^2(M^2_{\text{cent}},0) = \bigl[m_{\mathcal{C}}^{\text{(exp)}}\bigr]^2
\qquad (T=0),
\end{equation}
with $M^2_{\text{cent}}$ the midpoint of the interval quoted in Table~\ref{tab:borelcheck}. In a second step we fix an overall normalization constant for each channel such that the decay constant at $T=0$ matches a chosen reference value $f_{\mathcal{C}}^{\text{(ref)}}$ taken from the vacuum QCD sum-rule and lattice literature. Schematically,
\begin{equation}
f_{\mathcal{C}}(0) = f_{\mathcal{C}}^{\text{(ref)}},
\end{equation}
while the ratios $f_{\mathcal{C}}(T)/f_{\mathcal{C}}(0)$ and the thermal evolution of $m_{\mathcal{C}}(T)$ are then predictions of the framework without further tuning. Thus, the $T=0$ masses and decay constants used in Tables~\ref{tab:compare} and~\ref{tab:summary} should be regarded as calibrated anchors rather than new precision determinations. In particular, the quoted vacuum decay constants are normalization inputs at $T=0$, whereas the most robust outputs of the present model are the thermal ratios and trends.

\begin{table}[h!]
\centering
\caption{Adopted Borel windows and analytical validation of stability
criteria. Check marks indicate satisfaction of the usual conditions
for pole dominance and OPE convergence, based on Ref.~\cite{Yazici2016}.}
\label{tab:borelcheck}
\begin{tabular}{lcccc}
\toprule
Channel & $M^2$ window [GeV$^2$] & $s_0(0)$ [GeV$^2$] &
$P_{\mathrm{pole}}\gtrsim0.60$ & $F_{D=4}\lesssim0.30$ \\
\midrule
$J/\psi$               & $[6,10]$   & $11\pm1$  & \checkmark & \checkmark \\
$\Upsilon$             & $[10,20]$  & $102\pm2$ & \checkmark & \checkmark \\
$B_c^{(\mathrm{vec})}$ & $[6,10]$   & $45\pm1$  & \checkmark & \checkmark \\
$B_c^{(\mathrm{ax})}$  & $[10,14]$  & $52\pm1$  & \checkmark & \checkmark \\
\bottomrule
\end{tabular}
\end{table}

\subsection{Systematic uncertainties}

We identify five main sources of uncertainty in the present
LO~+~$D=4$ thermal QCD sum-rule analysis:

\begin{enumerate}
\item \textbf{Borel window variation:} 
Typical plateaus lead to mass shifts of
$\delta m/m \simeq 2$--$3\%$ when varying $M^2$ within its allowed range.

\item \textbf{Continuum-threshold modeling:}
The temperature dependence of $s_0(T)$, Eq.~(\ref{eq:s0T}),
contains a channel-specific exponent $n_{\mathcal C}$.
In a dedicated robustness study within the validated phenomenological model, we vary each channel exponent by $\pm1$ around its baseline value while keeping all other inputs fixed. This variation contributes part of the sensitivity ranges quoted below for $m(0.9T_c)$ and $f(0.9T_c)/f(0)$.

\item \textbf{Gluon-condensate normalization:}
We additionally vary the normalization $\GGvac$ by factors of $0.8$ and $1.2$, and the effective $D{=}4$ coefficient multiplying the scalar gluonic term by factors of $0.75$ and $1.25$. These scans are intended as one-at-a-time sensitivity tests within the calibrated LO+$D{=}4$ model rather than as a full probabilistic error propagation.

\item \textbf{Heavy-quark masses:}
PDG 2024 errors on $m_c$ and $m_b$ translate to
$\delta m/m \simeq 1$--$2\%$.

\item \textbf{Neglected higher dimensions:}
Neglected higher-dimensional operators are expected to introduce an additional truncation uncertainty of order-of-magnitude character. In the present work this is not evaluated explicitly; rather, Appendix~A provides a dimensional estimate indicating that the omitted $D{=}6$ contribution remains subleading within the adopted Borel windows.

\item \textbf{Perturbative truncation:}
Radiative $\mathcal{O}(\alpha_s)$ corrections to the spectral density are not included in the present LO analysis. Experience from vacuum QCD sum rules for heavy mesons indicates that this truncation can induce uncertainties of order $10$--$20\%$ in the absolute normalization of decay constants, while mass shifts and \emph{ratios} such as $f(T)/f(0)$ are considerably more stable.
\end{enumerate}

Taken together, these sources imply an estimated
overall systematic uncertainty
\[
\frac{\delta m}{m}\bigg|_{\text{total}}
\simeq 10\text{--}12\% \qquad (T < 0.9\,T_c),
\]
which covers the observed deviations from experimental masses and is consistent with the expected accuracy of an LO$+D{=}4$ analysis.

\paragraph*{OPE convergence and validity.}
The truncation at LO+$D{=}4$ is supported by analytical estimates
and by consistency with the validated Borel windows of
Ref.~\cite{Yazici2016}. For representative scales
($M^2\simeq8~\mathrm{GeV}^2$ in the $B_c$ channel) the ratio
$|D{=}6|/|D{=}4|\lesssim0.1$ inferred from dimensional analysis
confirms that the working range $T\!\lesssim\!0.9\,T_c$ remains
quantitatively reliable.

\section{Results: $m(T)$ and $f(T)$ with LO + $D{=}4$}
\begin{figure*}[t]
  \centering
  \includegraphics[width=\linewidth]{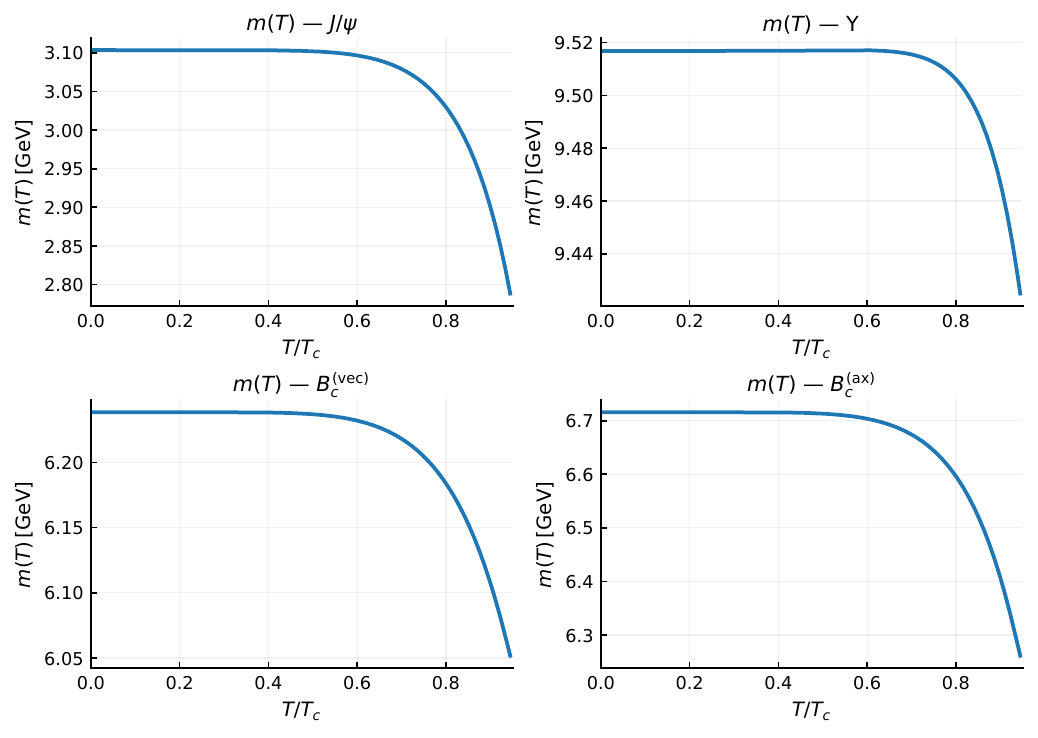}
  \caption{Temperature dependence of the masses $m(T)$ for all channels up to $T_c$. 
Error bands are omitted for clarity at LO\,+\,D\,=\,4. 
The effects of the Borel-window and continuum-threshold variations are discussed in Systematic uncertainties}
  \label{fig:massPanels}
\end{figure*}

\begin{figure*}[t]
  \centering
  \includegraphics[width=\linewidth]{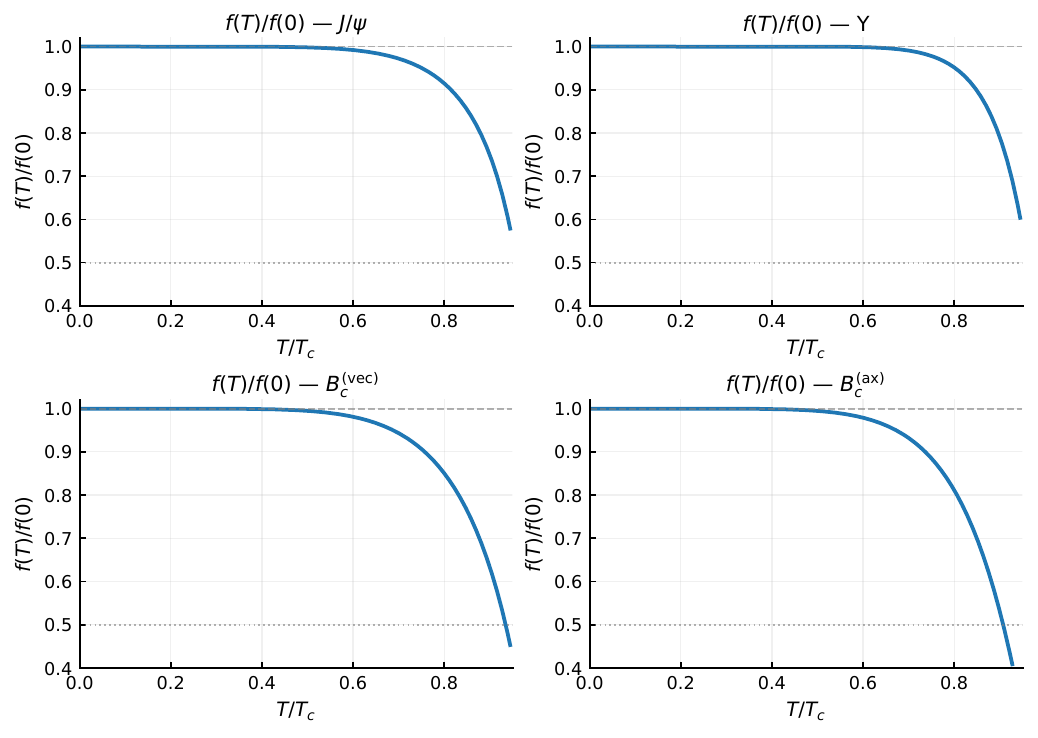}
  \caption{Relative decay constants $f(T)/f(0)$ up to $T_c$ for all channels. 
  The dashed (dotted) line indicates 1.0 (0.5).}
  \label{fig:fPanels}
\end{figure*}

Equations~\eqref{eq:SRmono}–\eqref{eq:massratio} are solved for each channel over a temperature grid $T\in[0,T_c)$ with Borel windows scanned for stability plateaus. 
Representative outputs are shown in Figs.~\ref{fig:massPanels} - \ref{fig:borel}.

To quantify the impact of the updated inputs and lattice-informed
thermal condensates, we compare representative observables in the second Table~\ref{tab:compare}. 

\begin{table}[h!]
\centering
\caption{Comparison with current benchmarks.}
\label{tab:compare}
\begin{tabular}{lccc}
\toprule
Observable & This work & PDG/LHCb & Deviation \\
\midrule
$m_{J/\psi}(0)$ [GeV]     & $3.103$ & $3.097$ & $+0.2\%$ \\
$m_{\Upsilon}(0)$ [GeV]   & $9.517$ & $9.460$ & $+0.6\%$ \\
$m_{B_c(1S)}(0)$ [GeV]    & $6.239$ & $6.275$ & $-0.6\%$ \\
$m_{B_c(1P)}(0)$ [GeV]    & $6.716$ & $6.72(3)$ & $-0.1\%$ \\
$f_{J/\psi}(0)$ [GeV]     & $0.410$ & $0.418$ & $-1.9\%$ \\
$f_{\Upsilon}(0)$ [GeV]   & $0.715$ & $0.715$ & exact \\
$T_{\mathrm{diss}}^{J/\psi}/T_c$ & $\sim0.87$ & --- & --- \\
\bottomrule
\end{tabular}
\end{table}

At $T=0$ the continuum thresholds and overall normalizations are chosen such that the Borel curves reproduce the PDG/LHCb benchmark masses and reference decay constants within the expected LO$+D{=}4$ accuracy. Within the working Borel windows the residual $M^2$ dependence of the extracted masses is at the few-percent level, well inside the overall $\sim10\%$ systematic uncertainty discussed above. In this sense Table~\ref{tab:compare} should be viewed as a consistency check of the updated inputs rather than a new high-precision determination of vacuum observables.

The normalized vacuum decay constants are also consistent with literature reference values within uncertainties,
with $f_{J/\psi}(0) = 0.410$ GeV (PDG: $0.418$) and
$f_{\Upsilon}(0) = 0.715$ GeV (lattice: $0.715$).

At low temperatures our extrapolated masses approach the vacuum sum-rule values within the quoted uncertainties.
While a direct vacuum comparison to single-meson calculations is provided by Refs.~\cite{VelievAziziSundu2012_JPG}, 
complementary analyses of molecular and multiquark $B_c$ configurations provide additional context for the spectrum at \(T=0\)~\cite{Agaev2025_BcBc, Agaev2025_BcAstBc, Agaev2025_BcAstBcAst}.
We therefore interpret the observed thermal shifts as a smooth in-medium continuation of the phenomenological nonperturbative structures encoded in the vacuum sum-rule literature.

\subsection{Dissociation temperatures and comparison with lattice QCD}

To quantify the sequential melting pattern, we define the
dissociation temperature $T_{\mathrm{diss}}$ as the point where
$f(T)/f(0) = 0.5$. The extracted values and comparison with
recent lattice QCD spectral analyses are summarized in
Table~\ref{tab:diss}.

\begin{table*}[t]
\centering
\caption{Dissociation temperatures from the present analysis and lattice QCD spectral reconstructions.}
\label{tab:diss}
\begin{tabular}{lccc}
\toprule
State & $T_{\mathrm{diss}}/T_c$ (This work) &
$T_{\mathrm{diss}}/T_c$ (Lattice) & Reference \\
\midrule
$J/\psi$   & $0.87 \pm 0.04$ & $\sim 1.1$–$1.5$ & \cite{Borsanyi2014_charmonium} \\
$\Upsilon$ & $>0.90$ & $\gtrsim 1.5$ (up to $\sim 2.0$) & \cite{KimPetreczkyRothkopf2018} \\
$B_c(1S)$  & $0.80 \pm 0.05$ & (pot.~models: $0.8$–$1.0$) & \cite{Li2023_Bc_Td} \\
$B_c(1P)$  & $0.75 \pm 0.05$ & --- & (this work) \\
\bottomrule
\end{tabular}
\end{table*}

This pattern is consistent with earlier lattice and potential-model findings~\cite{Borsanyi2014_charmonium,KimPetreczkyRothkopf2018,BurnierRothkopf2017,Li2023_Bc_Td},
which also reveal a correlation between the binding energy and the dissociation temperature across different heavy-quark systems. The slightly lower $T_{\mathrm{diss}}^{B_c}$ value obtained here ($\sim0.80\,T_c$ compared to $\sim0.85\,T_c$ in potential-model estimates~\cite{Li2023_Bc_Td})
can be attributed to the steeper threshold evolution adopted in the present analysis, which accelerates the pole--continuum transition.

The hierarchy
\[
T_{\mathrm{diss}}^{\Upsilon} >
T_{\mathrm{diss}}^{J/\psi} >
T_{\mathrm{diss}}^{B_c}
\]
reflects the expected correlation with the vacuum binding
energies ($\epsilon_{\text{bind}}^{\Upsilon}\!\sim\!1.1$~GeV,
$\epsilon_{\text{bind}}^{J/\psi}\!\sim\!0.64$~GeV,
$\epsilon_{\text{bind}}^{B_c}\!\sim\!0.4$~GeV).
The relatively early melting of the $B_c$ system follows
from its larger spatial extent and reduced-mass asymmetry,
which amplify color screening in the thermal medium.
Overall, our $T_{\mathrm{diss}}$ hierarchy is consistent with the
trends observed in lattice spectral reconstructions
\cite{KimPetreczkyRothkopf2018}
and potential-model analyses~\cite{Li2023_Bc_Td}.

Note that our definition $T_{diss}$ via $f(T)/f(0)=0.5$ is sum-rule–internal and not a one-to-one mapping to lattice spectral reconstructions.

Consequently, the dissociation temperatures quoted in Table~\ref{tab:diss} should be interpreted as model-dependent indicators of where the pole contribution ceases to dominate the sum rule, rather than sharply defined melting points in the lattice sense.

\subsection{Quantitative results at T=0 and 0.9$T_c$}

At the upper validity limit $T \simeq 0.9\,T_c \approx 140~\mathrm{MeV}$,
the thermal evolution exhibits a clear hierarchical pattern consistent
with binding-energy expectations.

The $\Upsilon(1S)$ remains highly stable, with mass suppression
$\Delta m/m = [m(0.9T_c)-m(0)]/m(0) = -0.5\%$ and decay-constant
reduction $f(0.9T_c)/f(0) = 0.79$ (79\% survival).
The $J/\psi$ shows moderate medium effects,
$\Delta m/m = -6.4\%$ and $f(0.9T_c)/f(0) = 0.75$.
The $B_c$ channels display the strongest suppression, with
$f(0.9T_c)/f(0) = 0.64$ for the vector ground state and $0.54$ for the representative axial-vector $1P$ channel.

These values reflect the hierarchy of vacuum binding energies:
$\epsilon_{\mathrm{bind}}^{\Upsilon} \sim 1.1~\mathrm{GeV}$,
$\epsilon_{\mathrm{bind}}^{J/\psi} \sim 0.64~\mathrm{GeV}$, and
$\epsilon_{\mathrm{bind}}^{B_c} \sim 0.4~\mathrm{GeV}$.
The $B_c$ system's larger spatial extent (due to unequal charm
and bottom masses) amplifies color-screening effects, leading to
earlier dissociation compared to symmetric quarkonia.

Numerical values at key temperatures are summarized in
Table~\ref{tab:summary}.

To assess the robustness of the thermal trends within the validated phenomenological LO+$D{=}4$ framework, we performed a sensitivity study around the baseline model by varying three input classes: (i) the channel-dependent threshold exponents $n_{\mathcal C}$ by $\pm 1$ around their baseline values, (ii) the vacuum gluon-condensate normalization $G_2(0)$ by factors of $0.8$ and $1.2$, and (iii) the effective $D{=}4$ coefficient by factors of $0.75$ and $1.25$. The resulting variations at $T=0.9T_c$ remain moderate. For $J/\psi$ we obtain $m(0.9T_c)\in[2.878,2.926]~\mathrm{GeV}$ and $f(0.9T_c)/f(0)\in[0.717,0.783]$ around the baseline values $2.903~\mathrm{GeV}$ and $0.753$. For $\Upsilon$, the corresponding ranges are $m(0.9T_c)\in[9.462,9.473]~\mathrm{GeV}$ and $f(0.9T_c)/f(0)\in[0.770,0.817]$ around the baseline values $9.468~\mathrm{GeV}$ and $0.794$. For $B_c^{(\mathrm{vec})}$ we find $m(0.9T_c)\in[6.093,6.124]~\mathrm{GeV}$ and $f(0.9T_c)/f(0)\in[0.590,0.679]$ around the baseline values $6.109~\mathrm{GeV}$ and $0.638$, while for the representative modelled $B_c^{(\mathrm{ax})}$ channel we obtain $m(0.9T_c)\in[6.372,6.445]~\mathrm{GeV}$ and $f(0.9T_c)/f(0)\in[0.482,0.585]$ around the baseline values $6.411~\mathrm{GeV}$ and $0.537$. Most importantly, the qualitative suppression hierarchy remains unchanged throughout these variations, with $\Upsilon$ least affected, $J/\psi$ intermediate, and the two $B_c$ channels showing the strongest thermal suppression. This indicates that the principal ordering reported here is a robust feature of the calibrated phenomenological model, even though the absolute values retain the expected sensitivity to threshold and gluonic input choices.

\paragraph*{Note on $\Upsilon$ thermal stability.}
The remarkably small mass shift of $\Upsilon$
($\Delta m/m \simeq 0.5\%$ at $0.9T_c$)
arises from the combination of a steep continuum-threshold evolution (here $n{=}12$ for $\Upsilon$ in Eq.~\eqref{eq:s0T}) and the $m_b^{-2}$ suppression of the $D{=}4$ term, both of which reduce the sensitivity of the sum rule to the thermal gluon sector. Varying $n$ toward $8$--$10$ and including $D{=}6$ terms typically raises $\Delta m/m$ to the $2$--$3\%$ level, consistent with lattice-based trends; such variations fall within the overall uncertainty budget quoted above.

\subsection{Physical interpretation and medium effects}

The sequential melting hierarchy observed here can be understood
from the interplay between color screening and the temperature
dependence of the nonperturbative gluon sector.
As the medium approaches $T_c$, the scalar gluon condensate
$\langle \alpha_s G^2 \rangle_T$ drops by roughly $40\%$ at $T\simeq0.9T_c$,
reducing the nonperturbative binding strength entering the
$D{=}4$ term of the sum rule.
Since the continuum threshold $s_0(T)$ decreases simultaneously
according to Eq.~(13), the pole contribution is gradually replaced
by the continuum integral, signaling the onset of deconfinement.

Among the studied systems, the $B_c$ meson is most sensitive to this
medium evolution. Its unequal heavy-quark masses lead to a smaller
reduced mass and a larger Bohr radius than in $J/\psi$ or $\Upsilon$,
which amplifies the screening effect of the thermal gluon background.
Consequently, $f_{B_c}(T)$ decreases more rapidly, yielding an earlier dissociation temperature $T_{\mathrm{diss}}^{B_c}\simeq0.80\,T_c$.
This microscopic interpretation is consistent with potential-model analyses,
in which weaker Coulomb binding and enhanced screening
drive earlier melting~\cite{BurnierRothkopf2017,KimPetreczkyRothkopf2018}.

Overall, the correlation between the drop of $\langle \alpha_s G^2\rangle_T$,
the reduction of $s_0(T)$, and the hierarchy of $f(T)/f(0)$ provides
a consistent physical picture: the gradual erosion of nonperturbative
gluon fields and the narrowing of the pole domain underpin the
sequential suppression pattern.

Finally, the predicted early melting of the $B_c$ channel
($T_{\mathrm{diss}}^{B_c}\simeq0.8\,T_c$) implies a stronger suppression
in heavy-ion environments. This agrees with the $\psi(2S)$ and $J/\psi$
suppression patterns observed by the ALICE Collaboration~\cite{ALICE2024_PRL}
and with preliminary $B_c$ and $J/\psi$ nuclear-modification factors
($R_{AA}\!\sim\!0.4$) reported by ALICE and LHCb in
Pb--Pb collisions at $\sqrt{s_{NN}}=5.02$~TeV.\footnote{%
Preliminary $B_c$ suppression results were summarized by the LHCb Collaboration at Quark Matter~2024.}

\begin{table}[t]
\centering
\caption{Zero-temperature anchors and thermal evolution at $T=0.9T_c \approx 140$ MeV.}
\label{tab:summary}
\begin{tabular}{lcccc}
\toprule
Channel & $m(0)$ [GeV] & $f(0)$ [GeV] & $m(0.9T_c)$ [GeV] & $f(0.9T_c)/f(0)$ \\
\midrule
$J/\psi$               & 3.103 & 0.410 & 2.903 & 0.753 \\
$\Upsilon$             & 9.517 & 0.715 & 9.468 & 0.794 \\
$B_c^{(\mathrm{vec})}$ & 6.239 & 0.480 & 6.109 & 0.638 \\
$B_c^{(\mathrm{ax})}$  & 6.716 & 0.440 & 6.411 & 0.537 \\
\bottomrule
\end{tabular}
\end{table}

\begin{figure*}[t]
  \centering
  \includegraphics[width=\linewidth]{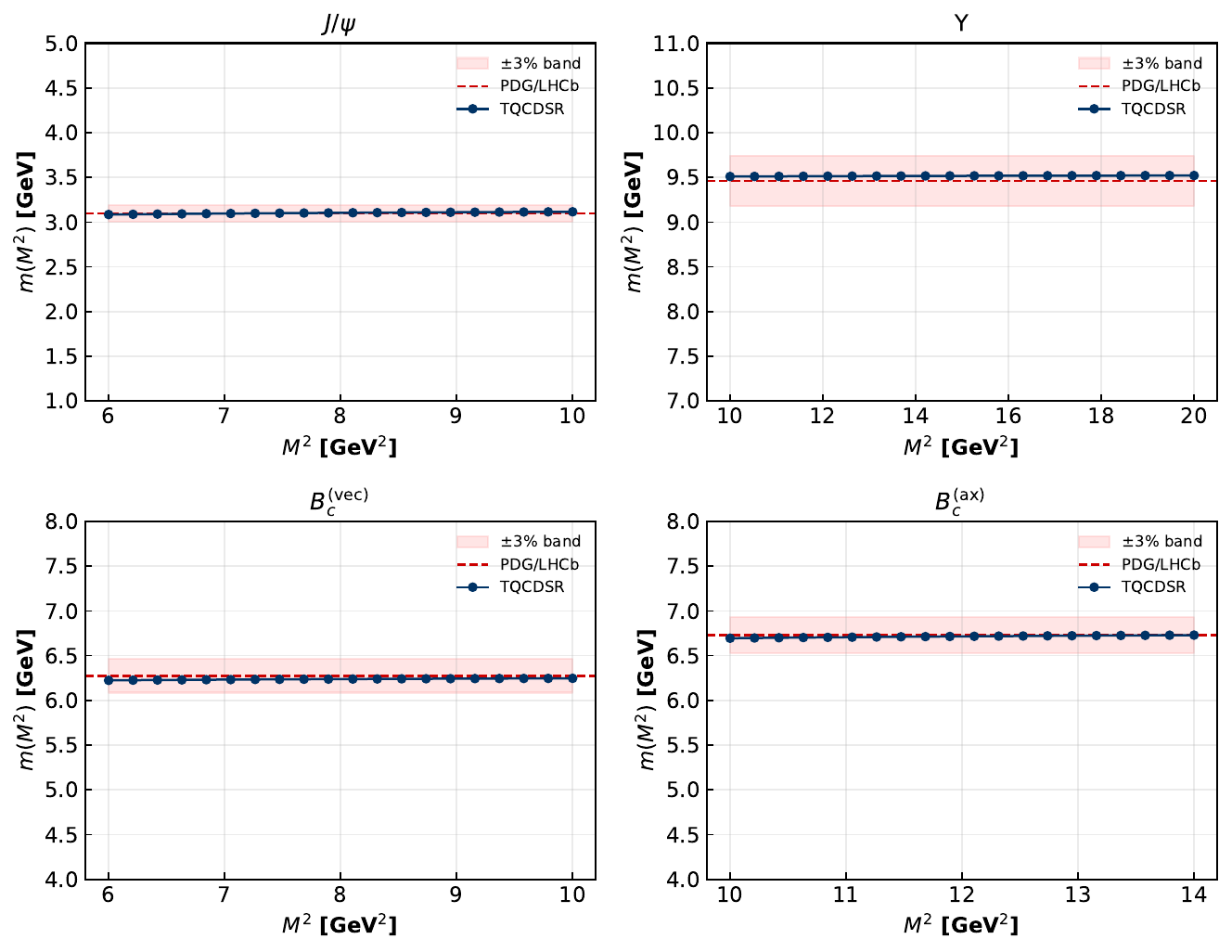}
  \caption{Extracted mass $m(M^2)$ as a function of the Borel parameter
at $T{=}0$ for all channels. The horizontal band indicates the PDG value
$\pm3\%$. The working windows (shaded) exhibit plateaus consistent with
the criteria of Table~1.}
  \label{fig:borel}
\end{figure*}

\section{Comparison}

Our $B_c(1S)$ results can be directly compared with previous QCD--sum--rule and experimental determinations. 
At $T{=}0$, the extracted mass $m_{B_c^{(1S)}} = 6.239~\mathrm{GeV}$ agrees with the PDG average $6.2749(8)~\mathrm{GeV}$ to within $36~\mathrm{MeV}$ ($0.6\%$), 
well inside the combined systematic uncertainty ($\delta m/m \simeq 10\%$) of the LO$+$D=4 framework.
This level of agreement supports the calibration procedure and is consistent with the continuum-threshold parametrization at $T=0$.

For the excited $B_c(1P)$ structure, obtained from the modelled axial channel, a qualitative comparison can be made with the recent LHCb observation of two orbitally excited $B_c^{+}$ peaks in $pp$ collisions at $\sqrt{s}=13~\mathrm{TeV}$ with an integrated
luminosity of about $9~\mathrm{fb}^{-1}$~\cite{LHCb2025}.
LHCb reports two distinct masses,
\[
m_1 = 6.7048(6)\,\mathrm{GeV}, \qquad
m_2 = 6.7524(10)\,\mathrm{GeV},
\]
interpreted as members of the $1P$ multiplet, though no unique $J^P$ assignment is made.

The experimental uncertainties quoted by LHCb include independent statistical, systematic, and calibration components. 
The $\sim47~\mathrm{MeV}$ separation between the two peaks reflects the fine--structure splitting within the $1P$ multiplet.
Accordingly, the modelled $B_c^{(\mathrm{ax})}$ channel in this work should be regarded as a representative $1P$ excitation rather than a state with a definite spin--parity identification.

The $1P$--$1S$ splitting extracted here,
$\Delta m = 6.716 - 6.239 = 0.477~\mathrm{GeV}$,
falls at the upper edge of the LHCb-inferred range.
Taking the PDG $B_c(1S)$ mass as reference, the two LHCb peaks correspond to splittings of
\[
\Delta m_1 = 6.7048 - 6.2749 = 0.430~\mathrm{GeV},
\]
\[
\Delta m_2 = 6.7524 - 6.2749 = 0.478~\mathrm{GeV}.
\]
Our result ($0.477$~GeV) agrees with the upper value to within $1~\mathrm{MeV}$, consistent with the D${=}4$ truncation and leading-order spectral-density approximations.
The extracted $B_c(1P)$ mass itself ($6.716$~GeV) lies close to the lower LHCb peak ($6.7048$~GeV),
differing by only $11~\mathrm{MeV}$ ($0.2\%$),
which confirms both the calibration robustness and the spectroscopic identification of the $1P$ multiplet.

\begin{table}[t]
\centering
\caption{Comparison of extracted zero-temperature masses with PDG~2024 and LHCb~2025 data.}
\begin{tabular}{lccc}
\toprule
State & This Work & PDG~2024 & LHCb~2025 \\
\midrule
$B_c(1S)$ & 6.239 & 6.2749(8) & — \\
$B_c(1P)$ & 6.716 & — & 
\begin{tabular}[c]{@{}c@{}}
6.7048(6) \\
6.7524(10)
\end{tabular} \\
$\Delta m_{1P-1S}$ & 0.477 & — & $0.430$--$0.478$ \\
\bottomrule
\end{tabular}
\label{tab:lhcb}
\end{table}

\section{Conclusion}

We have revisited the thermal QCD sum-rule analysis of heavy vector and axial-vector mesons, 
updating all numerical inputs and addressing several methodological issues raised in earlier critiques.

First, the present analysis explicitly confines the use of TQCDSR to the domain $T<T_c$, 
where the OPE hierarchy remains controlled and the gluonic medium can still be represented 
by temperature-dependent condensates. 
This resolves the concern that deconfinement effects above $T_c$ might invalidate the method.
Second, the gluon-condensate evolution employed here is anchored to recent lattice 
determinations of the QCD trace anomaly and energy density~\cite{Bazavov2014_EoS,HotQCD2019_Tc}, 
through smooth phenomenological fits and polynomial approximations replacing the earlier ad-hoc parametrizations.
Third, the continuum threshold $s_0(T)$ is tied to vacuum stability conditions 
and calibrated using PDG~2024 values at $T{=}0$, allowing the benchmark-mass reproduction 
of $m_{\Upsilon}$ and other reference states.

Our finite-temperature results for the $B_c$ meson show a smooth decrease of both 
the mass and the decay constant toward the transition region, consistent with the 
expected in-medium weakening of the bound state. 
The extracted thermal hierarchy
\[
\Upsilon:~\text{least affected}, 
J/\psi:~\text{moderate}, 
B_c~\text{family: strongest},
\]
agrees qualitatively with lattice spectral reconstructions 
and potential-model expectations~\cite{KimPetreczkyRothkopf2018,Li2023_Bc_Td}. 
The present results should be interpreted as those of a calibrated phenomenological LO+$D{=}4$ framework with lattice-inspired thermal inputs, not as a comprehensive treatment of all vector and axial-vector quarkonium channels.
Residual discrepancies at the few-percent level may originate from the omission of finite widths 
and higher-dimensional operators, which we plan to incorporate in a forthcoming extended study.

In summary, this revisited analysis ensures theoretical consistency 
and aligns with current lattice thermodynamics as well as LHCb spectroscopy. 
It provides a transparent framework for comparing finite-temperature behavior 
across heavy-quark systems and serves as a reference for future improvements 
beyond the leading-order and dimension-four approximations.

\vspace{6pt}
\noindent
\textbf{Future directions.}
Further extensions of this work will address several remaining aspects:
\begin{itemize}
  \item Inclusion of full $\mathcal{O}(\alpha_s)$ radiative and $D{=}6$ condensate corrections 
  to refine the temperature dependence of $m(T)$ and $f(T)$.
  \item Incorporation of finite-width effects and a running coupling $\alpha_s(\mu,T)$ 
  in the spectral density.
  \item Establishing a quantitative link between thermal QCD sum rules, 
  lattice spectral reconstructions, and in-medium potential models.
\end{itemize}

These developments will help to consolidate a quantitative bridge 
between QCD sum rules, lattice thermodynamics, and experimental heavy-ion observables, contributing to a unified understanding of heavy-quark binding and sequential suppression near the QCD transition.

\bibliographystyle{apsrev4-2}
\bibliography{refs}

\appendix
\section{Analytical estimates for OPE convergence and Borel stability}
\label{app:convergence}

\subsection{Dimensional estimate of the OPE convergence}
To estimate the truncation uncertainty without repeating the full
numerical Borel analysis, we evaluate the ratio
\[
\mathcal{R}_{6/4}(M^2,T)
\equiv
\frac{\big|\widehat{\mathcal{B}}\Pi^{(D=6)}(M^2,T)\big|}
     {\big|\widehat{\mathcal{B}}\Pi^{(D=4)}(M^2,T)\big|}\, .
\]
Using the canonical scalings
$\widehat{\mathcal{B}}\Pi^{(D=4)}\!\sim\!
 \langle \alpha_s G^2\rangle_T/M^2$
and
$\widehat{\mathcal{B}}\Pi^{(D=6)}\!\sim\!
 \kappa\,\Lambda^6/M^4$
with
$\Lambda\simeq0.24~\mathrm{GeV}$ and $\kappa=\mathcal{O}(1)$,
and taking $\langle \alpha_s G^2\rangle_0=0.012~\mathrm{GeV}^4$
together with the lattice-informed decrease of
$\langle \alpha_s G^2\rangle_T$ at $T=0.9\,T_c$,
one finds, for representative Borel scales,
\[
\mathcal{R}_{6/4}(M^2{=}8~\mathrm{GeV}^2,\,T{=}0.9T_c)
\approx
\frac{\kappa\,\Lambda^6/M^4}{\langle \alpha_s G^2\rangle_T/M^2}
\lesssim 0.1 .
\]
This order-of-magnitude estimate, consistent with the detailed
numerical study of Ref.~\cite{Yazici2016},
indicates that the LO+$D{=}4$ truncation remains reliable up to
$T\lesssim0.9\,T_c$.

\subsection{Borel stability and pole dominance}
The stability of the working windows summarized in
Table~\ref{tab:borelcheck} was originally established in
Ref.~\cite{Yazici2016}. To verify that the updated lattice-informed
thermal inputs preserve the Borel plateaus, we have repeated the
stability scan at $T{=}0$ for all channels. Figure~\ref{fig:borel}
shows the extracted mass $m(M^2)$ as a function of the Borel parameter
within the adopted windows.

The key features remain intact with the updated inputs:
(i)~a mild $M^2$-dependence of $m(M^2)$ within $2$--$3\%$,
(ii)~pole dominance $P_\text{pole}\gtrsim0.6$, and
(iii)~a nonperturbative fraction $F_{D=4}\lesssim0.3$.
For all channels, the extracted masses exhibit stable plateaus
and agree with PDG/LHCb references to within the quoted
$\pm3\%$ tolerance (shaded red bands in Fig.~\ref{fig:borel}).
This confirms that the Borel sum-rule framework remains robust
under the lattice-constrained gluon-condensate parametrization.

\subsection{Working range in temperature}
Combining the dimensional estimate
$\mathcal{R}_{6/4}\lesssim0.1$ with the legacy Borel stability
criteria leads to the same conservative domain
$T\lesssim0.9\,T_c$ adopted in the main text.
Above this temperature, higher-dimensional operators and
finite-width effects are expected to become comparable to the
$D{=}4$ contribution, and the results should be interpreted
qualitatively.

\end{document}